\documentclass[pra,twocolumn,amsmath,amssymb]{revtex4-1}

\usepackage{graphicx}
\usepackage{bm}
\usepackage{float}
\usepackage{setspace}
\usepackage{hyperref}
\usepackage{subfigure}
\usepackage{rotating}

\begin{document}
\title{Diffraction grating characterisation for cold-atom experiments}

\author{J.\ P.\ McGilligan, P.\ F.\ Griffin, E.\ Riis and A.\ S.\ Arnold $^{\footnote{aidan.arnold@strath.ac.uk}}$}

\affiliation{Dept. of Physics, SUPA, University of Strathclyde, Glasgow G4 0NG, UK}


\begin{abstract}
We have studied the optical properties of gratings micro-fabricated into semiconductor wafers, which can be used for simplifying cold-atom experiments. The study entailed characterisation of diffraction efficiency as a function of coating, periodicity, duty cycle and geometry using over 100 distinct gratings. The critical parameters of experimental use, such as diffraction angle and wavelength are also discussed, with an outlook to achieving optimal ultracold experimental conditions.
\end{abstract}

\maketitle

\section{Introduction}

Cold atom technologies have dominated precision measurements in recent years \cite{katori,deutsch,poli,gross}. The preference for cold atoms arises from the increased interrogation time that is provided in an isolated environment, allowing higher precision to be taken from a measurement \cite{bize}. Although many metrological experiments benefit from cold atom measurements \cite{ye, arnold}, the standard apparatus required is typically too large for portable devices. Despite the fact that current miniaturised metrological devices have proven highly successful \cite{kitching}, their precision is limited by their use of thermal atoms.

The source of cold atoms in most experiments is a standard magneto-optical trap (MOT) \cite{chu, monroe} which utilise 6 independent beams, each with their own alignment and polarisation optics. We have previously demonstrated a device that collects cold atoms in an optically compact geometry using a grating magneto-optical trap, GMOT \cite{vangeleyn2, nshii}, which is an extension of the equivalent MOT using a tetrahedral reflector \cite{vangeleyn1}. The simple design of the GMOT reduces the standard 6-beam MOT experimental set-up to one incident beam upon a surface-etched, silicon wafer diffraction grating. The grating uses the incident light and first diffracted orders to produce balanced radiation pressure, allowing us to trap a large number of atoms at sub-Doppler temperatures \cite{nshii,mcgilligan}. This greatly reduces the scale and complexity of optics used in laser cooling apparatus to facilitate applications \cite{himsworth, rolston, rosenbusch}.

In this paper we look to introduce a simple diffraction theory to assist the optical characterisation of these micro-fabricated diffraction gratings, with a view to aiding future cold atom quantum technologies. This study will be aimed towards an understanding of how metal coatings, periodicity, duty cycles and geometry affect the diffraction efficiency, a crucial parameter for creating balanced radiation pressure. Finally, we discuss additional parameters that have proven critical during our studies.

\section{Theory: Diffraction gratings}

Previous efforts made towards producing a reliable theory of diffraction require a typically complex derivation of Maxwell's equations \cite{golub, loewen}. However, here we look to introduce a simplified, phase based theory for determination of the first and zeroth order diffraction efficiencies.

\begin{figure*}[!t]
\centering
\includegraphics[width=12cm]{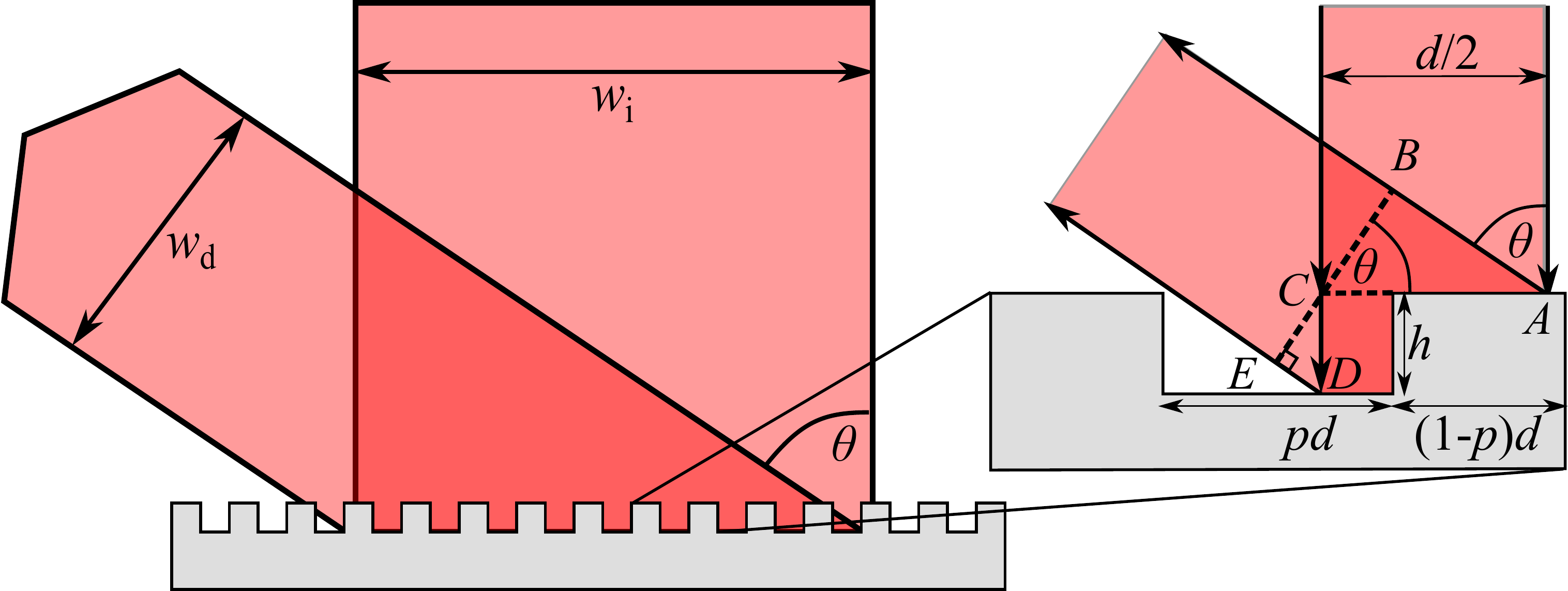}
\caption{Surface of a binary diffraction grating of etch depth
\ensuremath{h}, diffraction angle \ensuremath{\theta}, period \ensuremath{d} and duty cycle \ensuremath{p} and \ensuremath{1-p} for trough and peak respectively.
}
\end{figure*}

The diffraction grating is composed of a combination of reflecting elements arranged in a periodic array, separated by a distance comparable to the wavelength of study, as seen in Fig.~1. The separation between the reflective elements are troughs etched into the substrate, which is directly analogous to transmissive classical slits \cite{newport}. 

When studying a diffraction grating of period $d$ with incident light of wavelength $\lambda$ at an angle $\alpha$ to the grating normal, then a diffracted order will be produced at angle $\theta$ determined by the grating equation, $m \lambda = d(\sin\alpha +\sin\theta)$, where $m$ is an integer representing the diffracted order concerned. For incident light perfectly perpendicular to the grating ($\alpha=0$), the grating equation simplifies to the Bragg condition,
\begin{equation}
m\lambda=d\sin\theta,
\label{bragg}
\end{equation}
where $\theta$ is now the angle of diffraction.

Fig.\ 1 shows how the total electric field can be represented as the sum of diffracted orders from the trough and peak of the grating, which are weighted by their relative sizes $p d$ and $(1-p)d,$ respectively and phase shifted by the path difference between $AB=d/2 \sin \theta \,\,(= m \lambda/2$ from Eq.~\ref{bragg}) and $CDE =h (1+\cos \theta)$, i.e.:
\begin{equation}
E_{\rm tot}\propto p+(1-p) \, \exp \left[i \pi (m - 2 h(1+\cos \theta)/\lambda)\right], 
\label{Efield}
\end{equation}
where $h$ is the etch depth, and $\lambda$ the wavelength of incident light.

Using this electric field, in the case of 50/50 duty cycle, $p=1-p=0.5$, and a 1D grating then the intensity efficiency, $\eta_1$, in the first diffracted orders can be calculated via
\begin{equation}
\eta_{1}= R \frac{\left|1+ \exp \left[i \pi (1 -2 h(1+\cos \theta)/\lambda)\right]\right|^2}{8},
\label{intensity1}
\end{equation}
where $R$ is the reflectivity of the coating metal used, and the equation is valid for the symmetric diffraction orders $m=\pm 1.$  

Eq.\ (\ref{intensity1}) now provides a simple relation between the intensity of the light diffracted in the first order relative to the period of the grating. A simple model could thus assume that, for perfect diffraction and no second order, the zeroth order can be described by,
\begin{equation} 
2 \eta_{1}+\eta_{0}=R,
\label{intensity2}
\end{equation} 
If the etch depth, $h$, is designed such that $h=\lambda_d /4$, where $\lambda_d$ is the design wavelength and $\cos\theta \approx 1$ then Eq. (\ref{intensity1}) simplifies further to,
\begin{equation}
\eta_{1}=R \frac{\Big(1+\exp(i \frac{\pi \lambda_B}{2\lambda})\Big)^2}{8}
\label{wavelength}
\end{equation}
To apply these first order diffraction efficiencies, $\eta_1$, to 2D gratings we simply multiply by 1/2, to account for twice as many diffracted beams.

To determine how these diffracted efficiencies relate to creating a balanced radiation pressure we must account for the vertical intensity balance between the incident, $I_{i}$, and the diffracted orders, $I_{\rm d}$, described as $\frac{I_{\rm d}}{I_{\rm i}}= \eta_{1} \frac{w_{\rm i}}{w_{\rm d}} = \frac{\eta_{1}}{\cos \theta} $, where $w_{\rm i}$ (Fig.~1) is the incident beam waist and $w_{\rm d}$ is the diffracted beam waist. The radial balance is not considered as this is automatic if the beam centre is positioned on the grating center. 
The net incident intensity on the grating $I_{\rm i} (1-\eta_0)$ is ideally balanced with the component of the diffracted intensity which is anti-parallel with the incident light, i.e. $N I_{\rm d} \cos \theta$ where $N$ accounts for the number of diffracted first orders, which simplifies to $N I_{\rm i} \eta_1$. Thus, the balance between incident and diffracted light, perpendicular to the grating and taking the zeroth order into account, is described mathematically through the dimensionless quantity,
\begin{equation}
\eta_{\rm B}=N\eta_1/(1-\eta_0)
\label{intensitybalance}
\end{equation}
which is ideally one.

\section{Experiment: Grating characterisation}
\begin{figure}[!t]
\centering
\begin{minipage}{.49\columnwidth}
\centering \includegraphics[width=3cm]{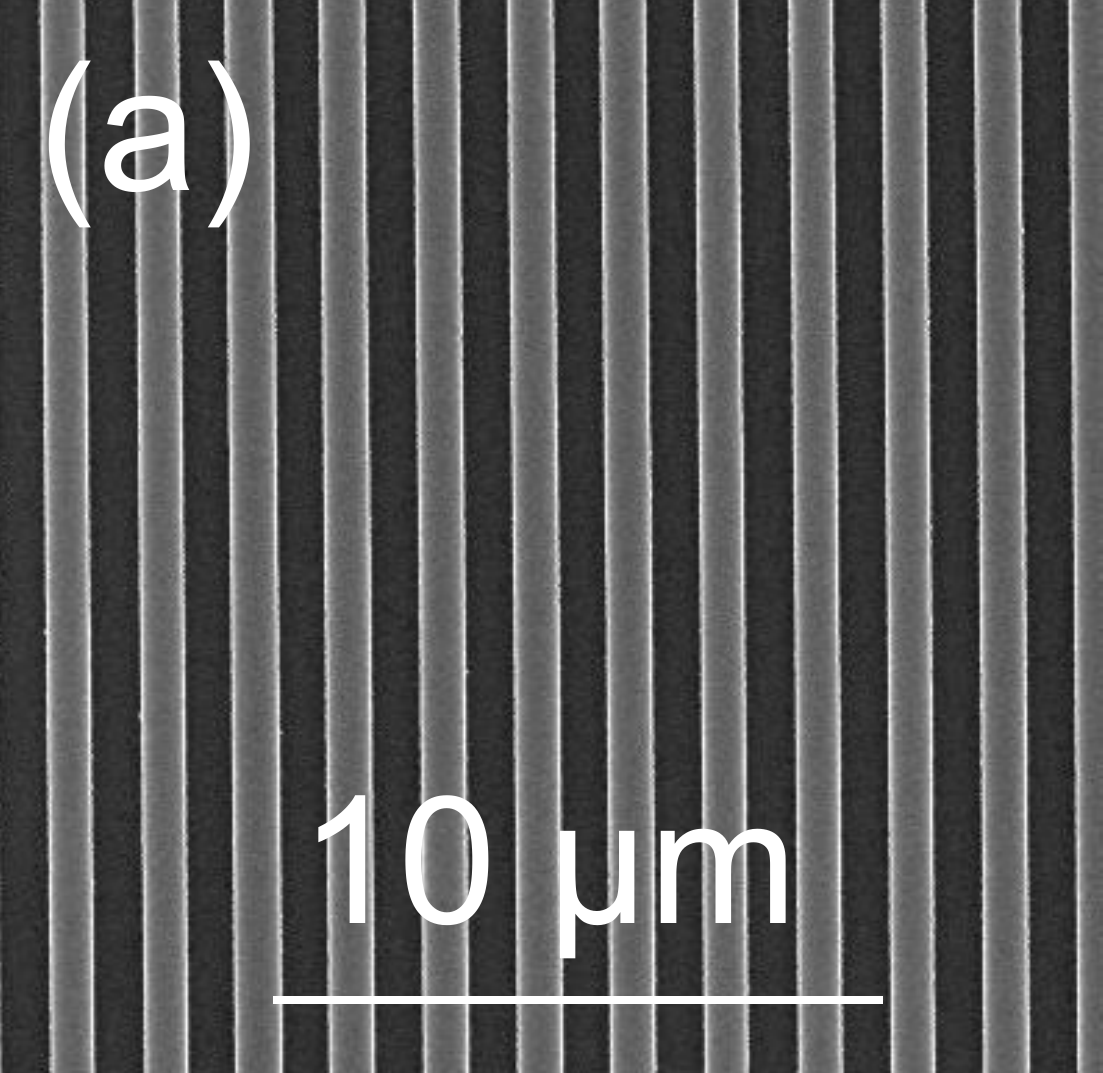}
\end{minipage}
\begin{minipage}{.49\columnwidth}
\centering \includegraphics[width=3cm]{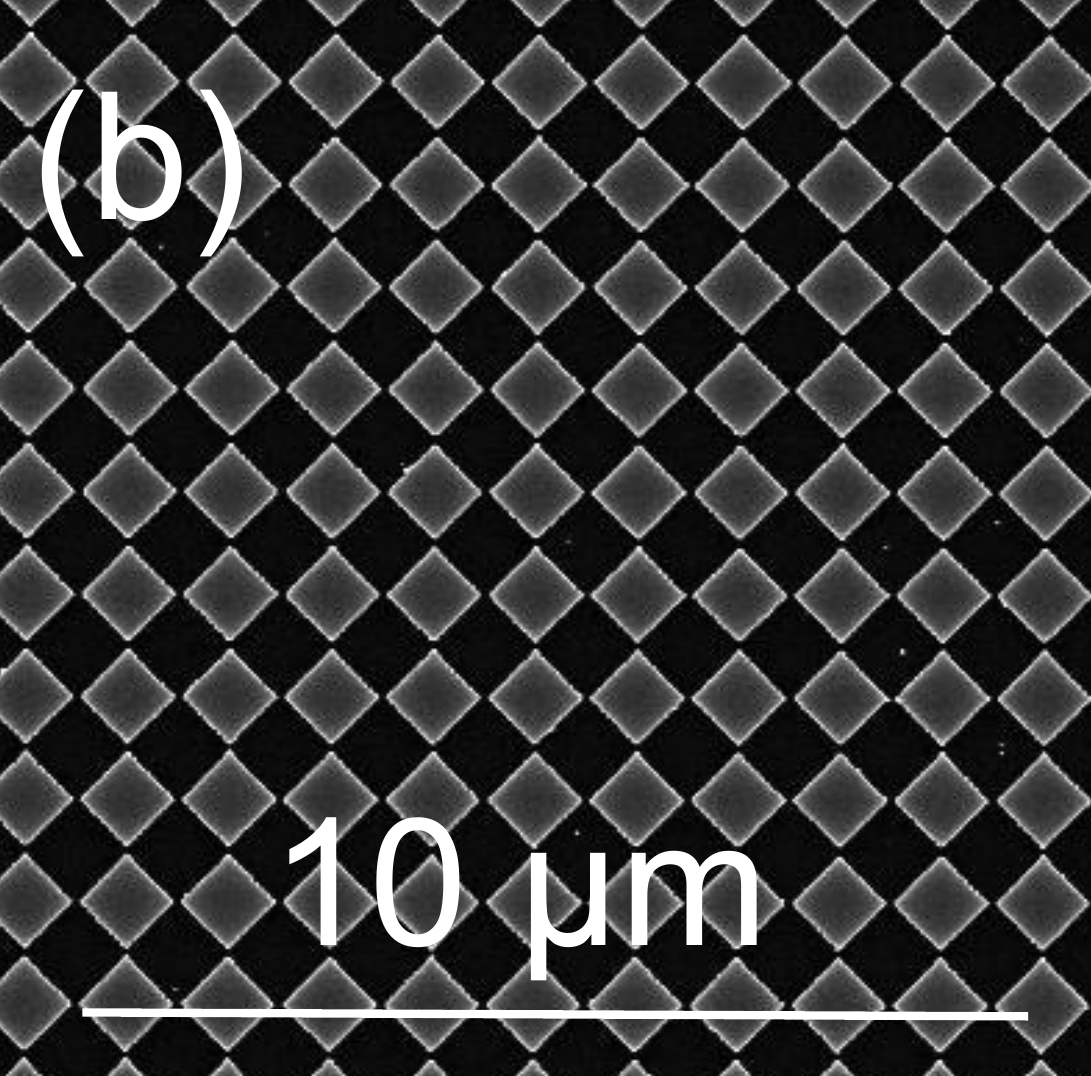}
\end{minipage}
\includegraphics[width=8cm]{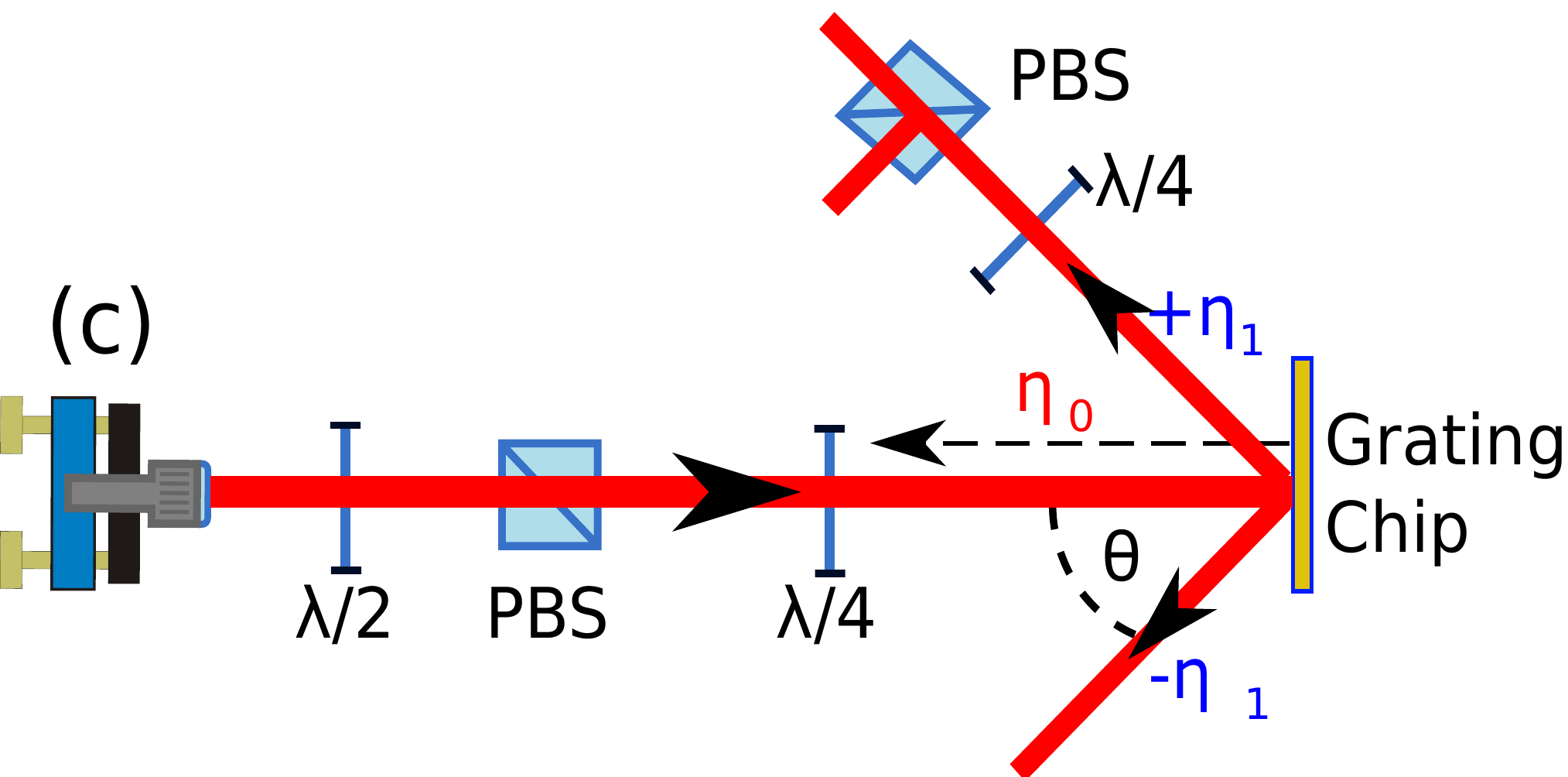}
\caption{(a) (b): Scanning electron microscope images of 1D and 2D gratings respectively. (c): Set-up used for grating efficiency and polarisation purity analysis. Abbreviations are $\lambda/2$ and $\lambda/4$ for the half and quarter wave-plates respectively, PBS for polarising beam splitter, $\eta_i$ represents the relative power in the $i^{\rm th}$ order of diffraction and $\theta$ is the angle of diffraction.}
\label{setup}
\end{figure}

The diffraction gratings used were manufactured with a dry etch into silicon wafers and patterned using electron beam lithography \cite{tseng,kley} to an ideal etch depth of $h=\lambda_B/4$ ($\lambda_B$=780~nm) and chosen periodicity. The wafer on which the Au gratings were etched is composed of silicon topped with 10~nm Ti and 20~nm Pt, whereas no adhesion layer was required for an Al grating. These are then sputter coated with a variable thickness coating layer. The geometry of the etch can vary between one dimensional, 1D, and two dimensional, 2D, gratings as illustrated in the scanning electron microscope images in Fig. \ref{setup} (a) and (b), respectively. The 2D grating produces four first order diffracted terms compared to the two produced in a 1D geometry.

To produce the ideal grating, a thorough investigation of how fabrication parameters affect the diffraction efficiency is required. The most logical way to determine the optimum settings for future diffraction gratings was to commission the construction of over one hundred 2~mm $\times$ 2~mm gratings, produced with a variety of periodicity, duty cycles, geometrical layout, coating metal and coating thickness. The best method to measure the properties of the large quantity of diffraction gratings was to construct a dedicated testing station with incident collimated, circularly polarised light of known wavelength and power, as can be seen in Fig. \ref{setup}.

\begin{figure}[!t]
\centering
\begin{minipage}{.49\columnwidth}
\includegraphics[width=4cm]{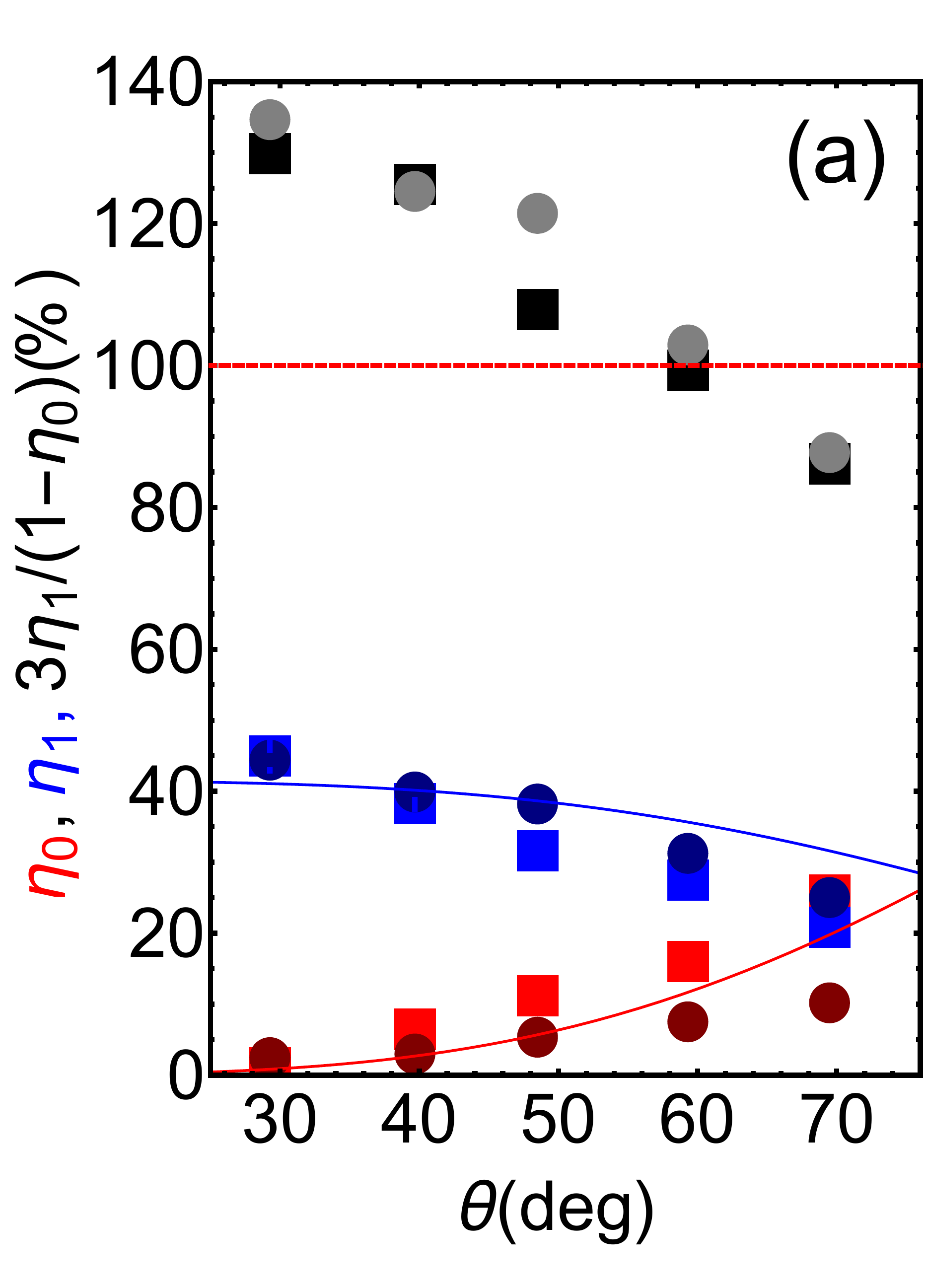}
\end{minipage}
\begin{minipage}{.49\columnwidth}
\includegraphics[width=4cm]{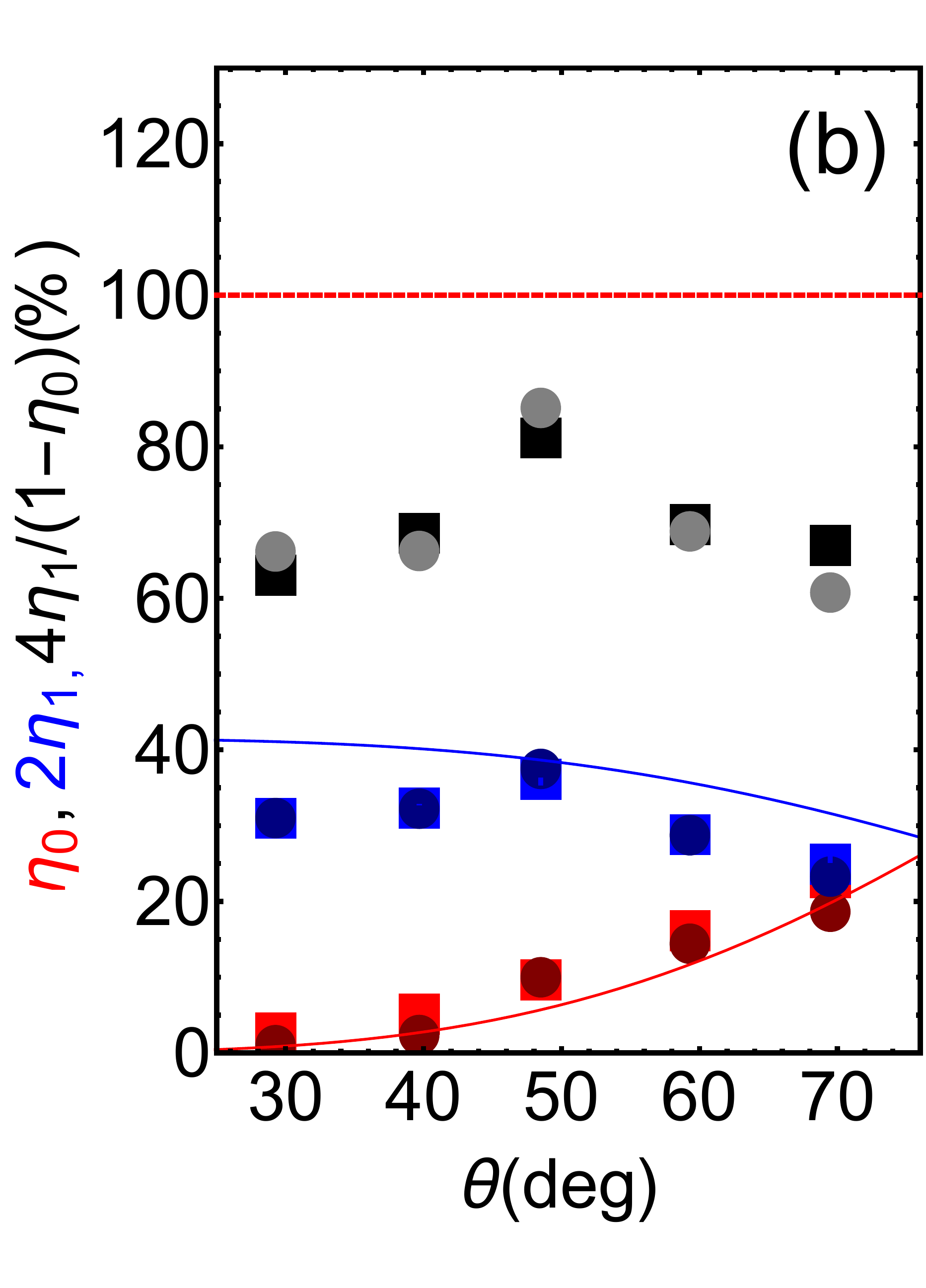}
\end{minipage}
\caption{Diffraction angle vs.\ radiation balance and diffraction efficiency. (a): 1D gratings with 80~nm Au coating. (b): 2D gratings with 80~nm Au coating. Blue and red represent the diffracted $(\eta_1)$ and reflected $(\eta_0)$ orders respectively, with black illustrating the radiation balance for gratings with duty cycles of 40$\%$:60$\%$ and 50$\%$:50$\%$ (circles and squares).}
\label{collective}
\end{figure}

\begin{figure}[!t]
\centering
\begin{minipage}{.49\columnwidth}
\includegraphics[width=4cm]{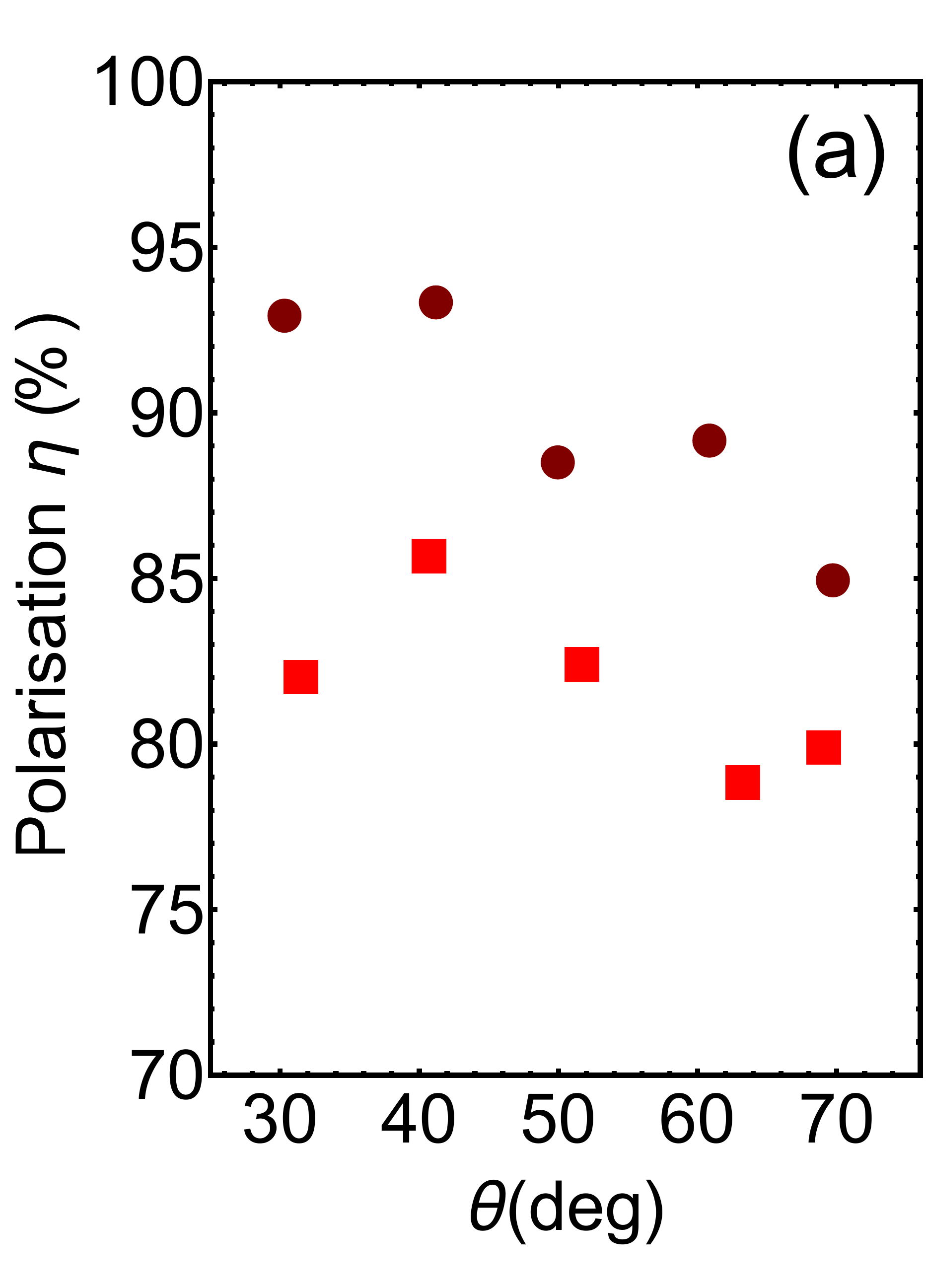}
\end{minipage}
\begin{minipage}{.49\columnwidth}
\includegraphics[width=4cm]{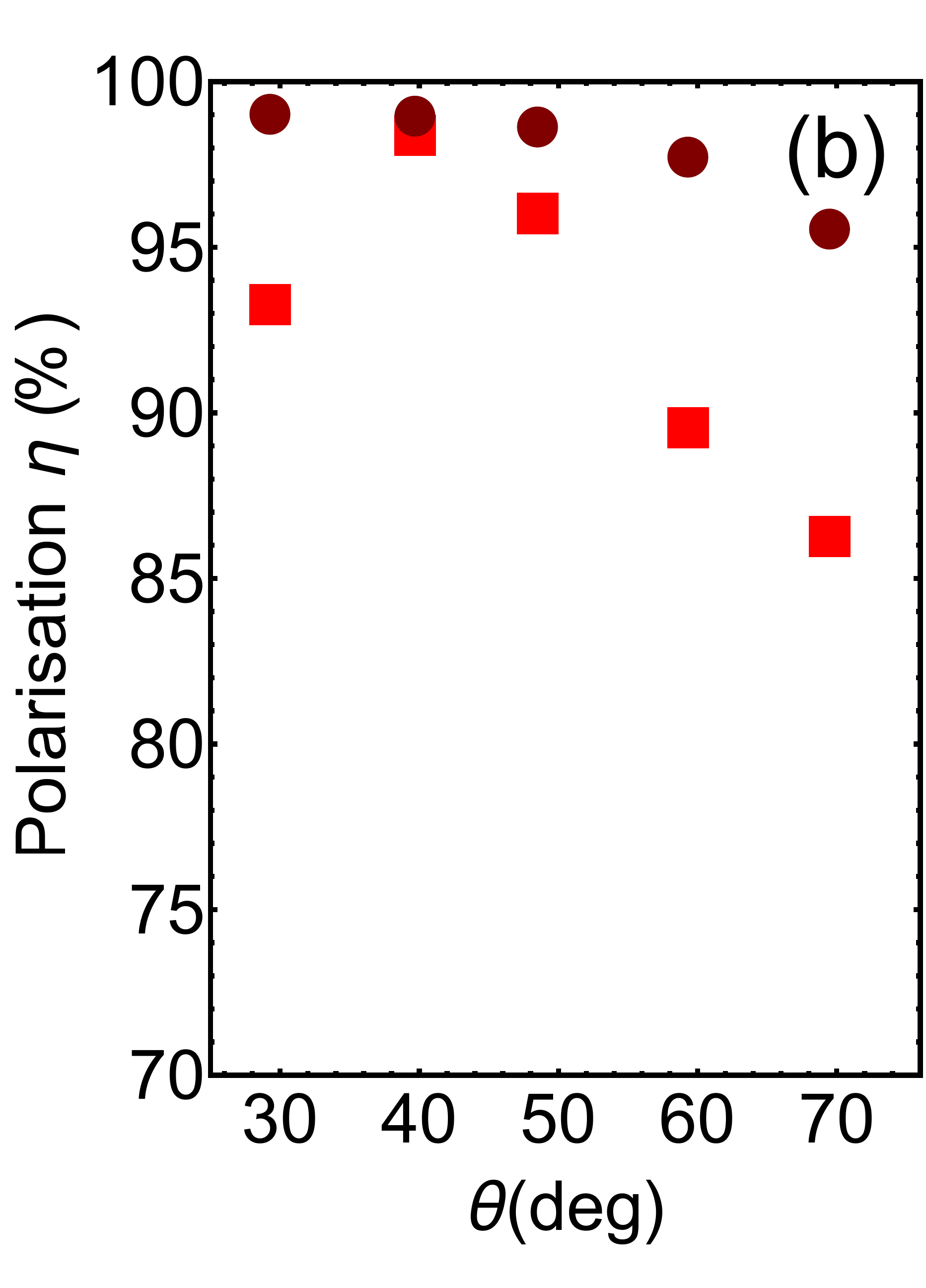}
\end{minipage}
\caption{Diffraction angle vs. polarisation purity. (a): 1D gratings with 80~nm Au and Alumina layer. (b): 1D gratings with 80~nm Au. Associated Appendix Fig.~\ref{pol}  provides detailed polarisation efficiency information for 1D and 2D gratings with two different thicknesses of gold and aluminium coating. Fig.~\ref{pol} also shows the effect of the 20nm alumina coating for three sets of 1D grating chips. In all images duty cycles of 40$\%$:60$\%$ and 50$\%$:50$\%$ are indicated by circles and squares, respectively.}
\label{polarisation}
\end{figure}

Once the grating was mounted in the set-up, the zeroth order was carefully aligned to ensure the incident light was perpendicular to the grating. The inherent need for this alignment will be discussed later. The position of the diffracted order was noted, and $\theta$ measured. This allowed the periodicity to be inferred through the Bragg condition, Eq.~(\ref{bragg}). The diffracted order is measured for diffracted power, then passes through a $\lambda/4$ plate and PBS to measure any degradation of polarisation that may have occurred during diffraction. The results of this investigation can be seen in Figs.~\ref{collective} and \ref{polarisation}, and in greater detail in their associated Appendix Figs.~\ref{diff} and \ref{pol} respectively. 

Fig. \ref{collective} depicts how the relative diffracted power and beam intensity balance vary with diffraction angle, $\theta$ for 1D and 2D gratings. The circles and squares represent gratings with spatial dimension etched:unetched duty cycles over one grating period of 40$\%$:60$\%$ and 50$\%$:50$\%$ respectively. The blue and red fits are derived from Eq. (\ref{intensity1}) and (\ref{intensity2}) for the first and zeroth diffracted order, where $R=0.75$ to account for the 98$\%$ reflectivity of gold and a loss mechanism found in the gratings, discussed later. Both data sets provided have a coating of 80~nm Au, however, further investigation was carried out into thicker coatings on Au as well as Al, with both 1D and 2D geometries. The results provided in Fig. \ref{collective} are typical of all data sets recorded, with any discrepancy discussed, however associated Appendix Figs.~\ref{diff} and \ref{pol} provide detailed diffraction efficiency and polarisation purity information, respectively, for 1D and 2D gratings with two different thicknesses of gold and aluminium coating. Moreover, Fig. \ref{diff} also shows that -- for the 1D gratings -- gold with a thin 20nm alumina coating has similar reflectivity to plain gold. The purpose of the alumina coating was to introduce a layer between the Au surface of the grating and the Rb metal vapour inside the vacuum system, which corrodes the Au.

The first point of interest is the decrease of the diffracted order relative to diffraction angle. As the first order decreases, the light in the zeroth order increases at the same rate, maintaining a close to constant total power. This decay is weaker in the gratings with 40$\%$:60$\%$ duty cycle, making this the preferable choice to 50$\%$:50$\%$ duty cycle. Analysis of experimental data proved that a thicker coating material causes no notable change in the 1D gratings. However, the diffraction efficiency was seen to increase by $\approx$ 10$\%$ when twice the coating thickness was applied to 2D gratings.  A gold coating produces a stronger diffracted order than that of the aluminium of similar coating depth due to the higher reflectivity of gold. The results from the duty cycle are conclusive that 40$\%$:60$\%$ duty cycle produces a lower reflected order and higher diffraction efficiency. The reasoning for this is not completely understood, but further modelling will be provided in Ref.~\cite{cotter}.

Fig. \ref{collective} also illustrates the balance of light force from Eq.~(\ref{intensitybalance}) for the respective geometry of the gratings intended use, as a function of diffraction angle. The dashed line at 100$\%$ represents axial balance between the incident downward beam and diffracted upward orders. This balancing force is notably higher in the 1D gratings compared to the 2D gratings as the 1D gratings only diffract into 2 beams rather than 4. However, with appropriate filtering of the incident beam, this can be overcome to produce well balanced radiation forces \cite{mcgilligan} required for laser cooling \cite{chu2, dalibard}. Using a 4 beam configuration with a linear grating provides close to ideal balance already without need for further adaptations to the apparatus. The results are typical of 1D and 2D gratings. Testing was also carried out on Au coated gratings with a top layer of alumina. Although there was no difference in diffraction efficiency between the gratings with and without the alumina, the additional layer was observed to degrade the polarisation purity of the diffracted order, Fig. \ref{polarisation}.

The polarisation purity $\eta$ refers to the ratio of correctly handed circular light (for MOT operation) to total light after the polarisation analyser PBS (Fig.~\ref{setup}c) in the first diffracted order. When measured against periodicity, this purity was typically above 95$\%$ for a 40$\%$:60$\%$ duty cycle (circles). The lower duty cycle of 50$\%$:50$\%$ (squares) consistently produced a weaker purity, which was noted to worsen in the case of an alumina coating. This side effect of using alumina coating could be mildly detrimental to experiments requiring in-vacuo gratings as the trapping force is proportional to $2\eta - 1$. \cite{vangeleyn2}.

\section{Experiment: Lost light}

\begin{figure}[!b]
\centering
\includegraphics[width=8.4cm]{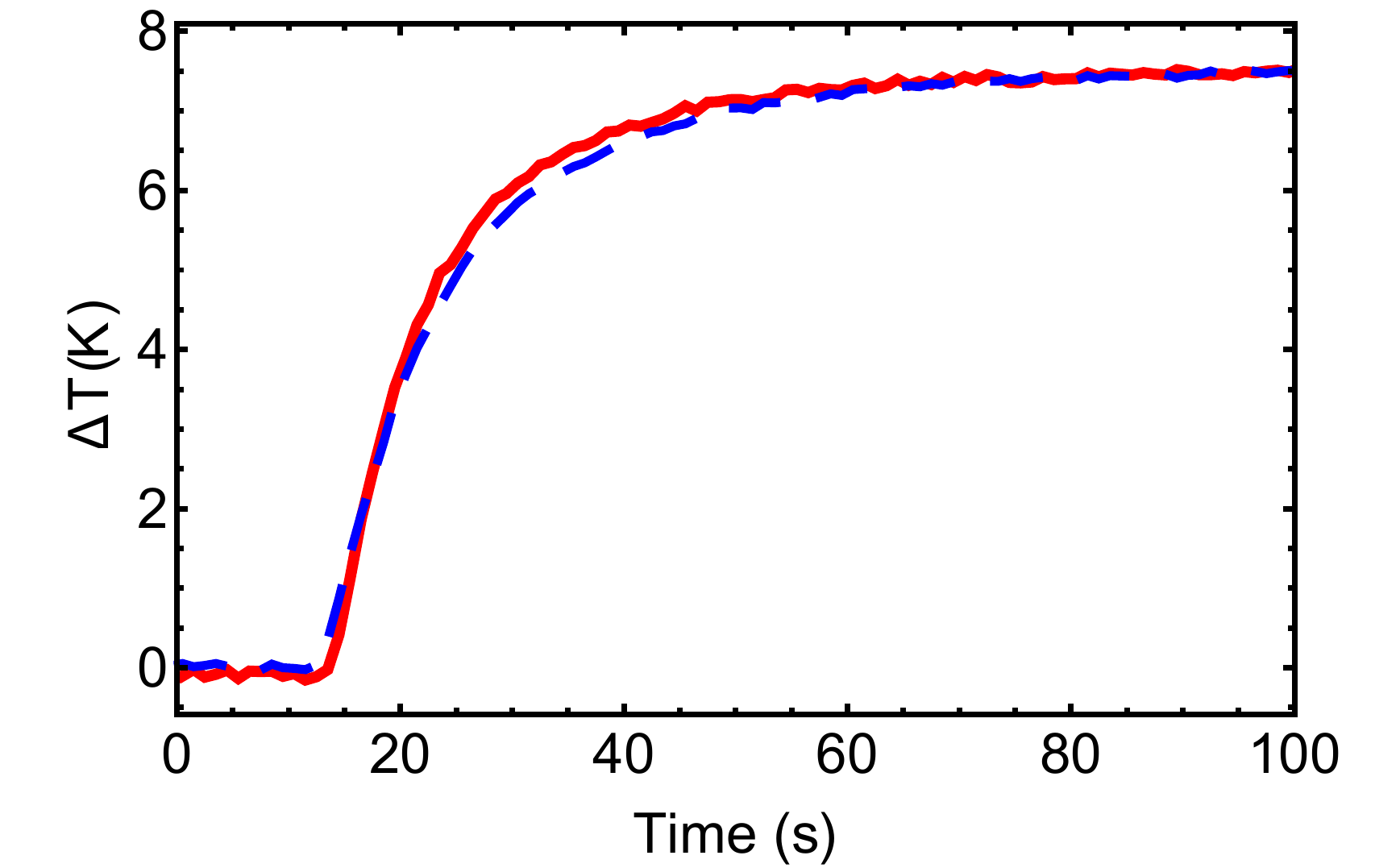}
\caption{Absorption rate of 1D Au coated grating heated with laser light (red) and calibrated with Ohmic heating (dashed blue).}
\label{heating}
\end{figure}

As has been pointed out with the diffraction efficiency data, the total power measured in the diffracted orders fell short of the incident power by $\approx$18~$\%$. The theoretical reasoning for this can be conceived as shadowing of the beam/diffraction losses in the pits of the grating and is discussed in more detail in \cite{cotter} -- here we discuss experiment measurement of the losses. An initial investigation into the elusive light was carried out through the investigation of the absorption profile of a grating by measuring the transfer of light to heat. For this, a small thermistor was well insulated to the back of a 4~mm $\times$ 4~mm Au coated diffraction grating, to read out the heating rate of the grating with a known incident laser power. This absorption rate can be seen in Fig. \ref{heating}.

To calibrate the grating heating to a known power, Ohmic heating was applied through a $1.5\,$k$\Omega$ resistor thermally attached to the back of the grating, separate from the thermistor. This resistor was connected in series to a voltage supply to deposit known amounts of power onto the grating, whilst measuring the heating rate. This Ohmic heating rate was then matched to that of the laser heating to determine the amount of laser power absorbed by the grating during the heating process. 

In order to account for thermal gradients in the area of the grating, the measurement procedure was also carried out for a plane Au coated wafer. Since plain Au has a known 3~$\%$ absorption at 780~nm \cite{NIST} we could use this to account for thermal gradients in the measurement area, that could then be applied to the grating data. Applying this correction factor results in 12$\pm$2~$\%$ of incident light being absorbed by the diffraction grating.

A further study into the possibility of the missing light being scattered was carried out to see if fabrication imperfections were projecting light into unwanted diffraction angles \cite{haring}. This was carried out by taking long exposure images around a 90$^{\circ}$ plane of diffraction and normalising the range of exposure times to determine the relative power in an minuscule peaks found. The data from this provided that $<$1$\%$ of lost light was being scattered by the grating.

\section{Experiment: Critical parameters}
\begin{figure*}[!t]
\centering
\includegraphics[width=13cm]{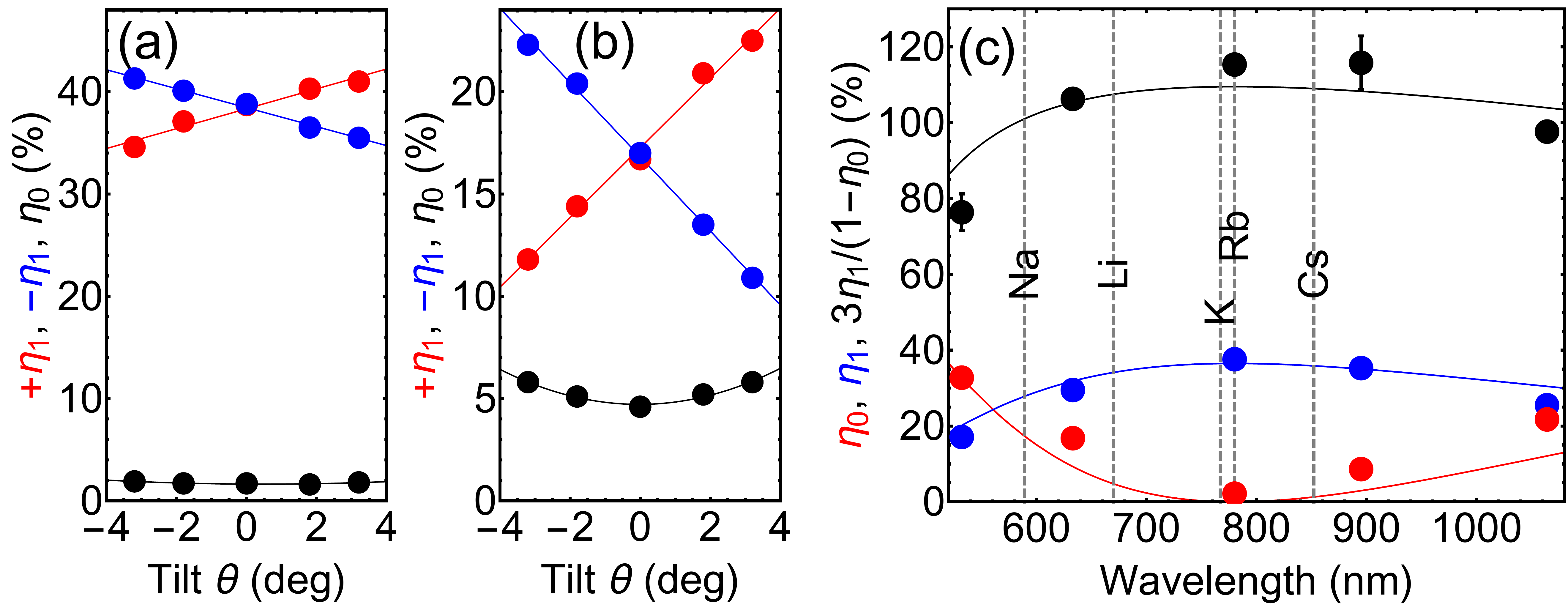}
\caption{Left: The grating angle tilt vs. the power in the relative diffracted orders with simple linear/parabolic fits for (a): Al 1D grating, d=1478~nm. (b): Au 2D grating, d=1056~nm. (c): Using the same grating as in (a), the wavelength of incident light is varied and recorded against the powers of first and zeroth diffracted orders and fit against theory from Eq.~(\ref{wavelength}). Black data points represent the intensity balance from Eq.~(\ref{intensitybalance}). The same set-up as in Fig. \ref{setup} was used, except the $\lambda/4$ wave-plates were replaced by Fresnel rhombs due to their achromatic retardance.}
\label{critical}
\end{figure*}

When implementing the diffraction grating into an experimental set-up, it is mounted perpendicular to the incident beam, however, the extent to which this angle of incidence can be varied is an important consideration. We investigated the angle sensitivity using the same set-up as in Fig. \ref{setup} (c), with variable tilt applied to the grating mount. Whilst in this configuration, a known amount of light was incident upon the grating, held at a variable tilt angle whilst the diffracted orders where measured. This procedure was carried out for both 1D and 2D gratings, the results of which are seen in Fig. \ref{critical} (a) - (b).

The blue and red data sets represent the opposite first diffracted orders, with black representing the zeroth, with best fitting lines and parabola applied. Fig. \ref{critical} (a)- (b) demonstrates that a small deviation from $90^{\circ}$ will symmetrically imbalance the first diffracted orders, and increase the unwanted zeroth order. This asymmetry vs angle is markedly more for 2D gratings (b) in comparison to 1D gratings (a).

It would also be of importance to know how the diffraction gratings' diffracted efficiency varies with the wavelength of incident light, as a wide bandwidth of wavelengths could unlock alkaline earth metals as possible species to be used in the grating MOT configuration. Additionally, knowing the dependence upon $\lambda$ would also provide understanding of etch depth, where $h=\lambda/4$. For this investigation the same set-up was used as in Fig. \ref{setup}, with 5 different lasers, covering a range of wavelengths seen in Fig. \ref{critical} (c). The red and blue data points depict the measurements of first and zeroth diffracted orders, with the fits derived from Eq.~(\ref{wavelength}). The black data points again depict the intensity balance from Eq.~(\ref{intensitybalance}). As is illustrated, the grating would deliver reasonably balanced cooling within $\pm 200\,$nm of the design wavelength of $780\,$nm.

\section{Conclusions and outlook}
In summary, we have presented our findings on producing next generation diffraction gratings for cold atom experiments. This study has illustrated the preferred fabrication parameters for optimising the grating diffraction efficiency and polarisation purity. 

We conclude that future gratings should be created with a higher duty cycle, as was seen from our study between 50$\%$:50$\%$ and 40$\%$:60$\%$ duty cycles. The study of coating thickness has also demonstrated that for the 2D geometry the thicker coating metal is preferable for higher diffraction efficiency. If an additional coating of alumina is placed on top of the grating for use within a vacuum system then a degradation of the polarisation purity has been noted. However, the efficiency of the weaker polarisation, with the correct duty cycle, does not hinder the creation of a MOT.

Finally, the critical parameters discussed demonstrate that, when implemented experimentally, the grating should be as close to perfectly perpendicular as possible to maintain balance between the diffracted orders, especially for the 2D gratings. The study of wavelength demonstrates broadband diffractive efficiency, opening the door to the cooling of elements on multiple atomic transitions.

\section{Funding}

EPSRC (EP/M013294/1); DSTL (DSTLX-100095636R); ESA (4000110231/13/NL/PA).

\section{Appendix: Characterising grating diffraction and polarisation effects vs.\ coating}
\begin{figure*}[!ph]
\includegraphics[width=1.8\columnwidth]{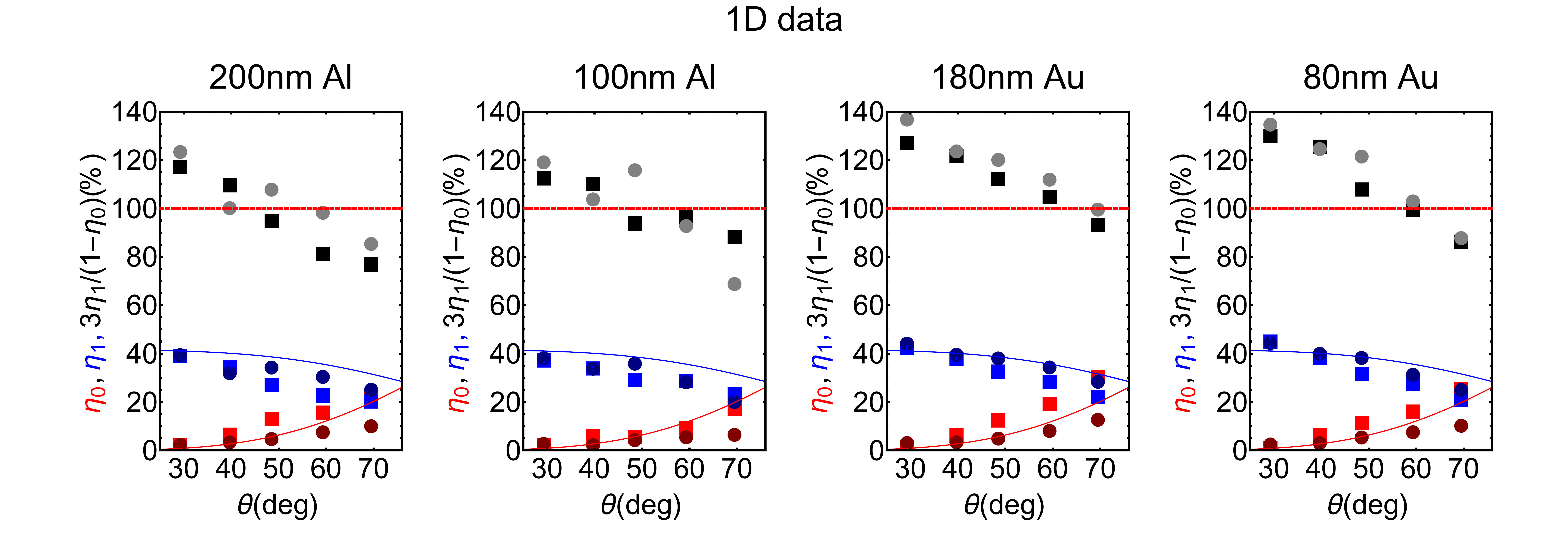}
\begin{minipage}{\columnwidth}
\vspace{2cm}
\end{minipage}
\centering \includegraphics[width=1.45 \columnwidth]{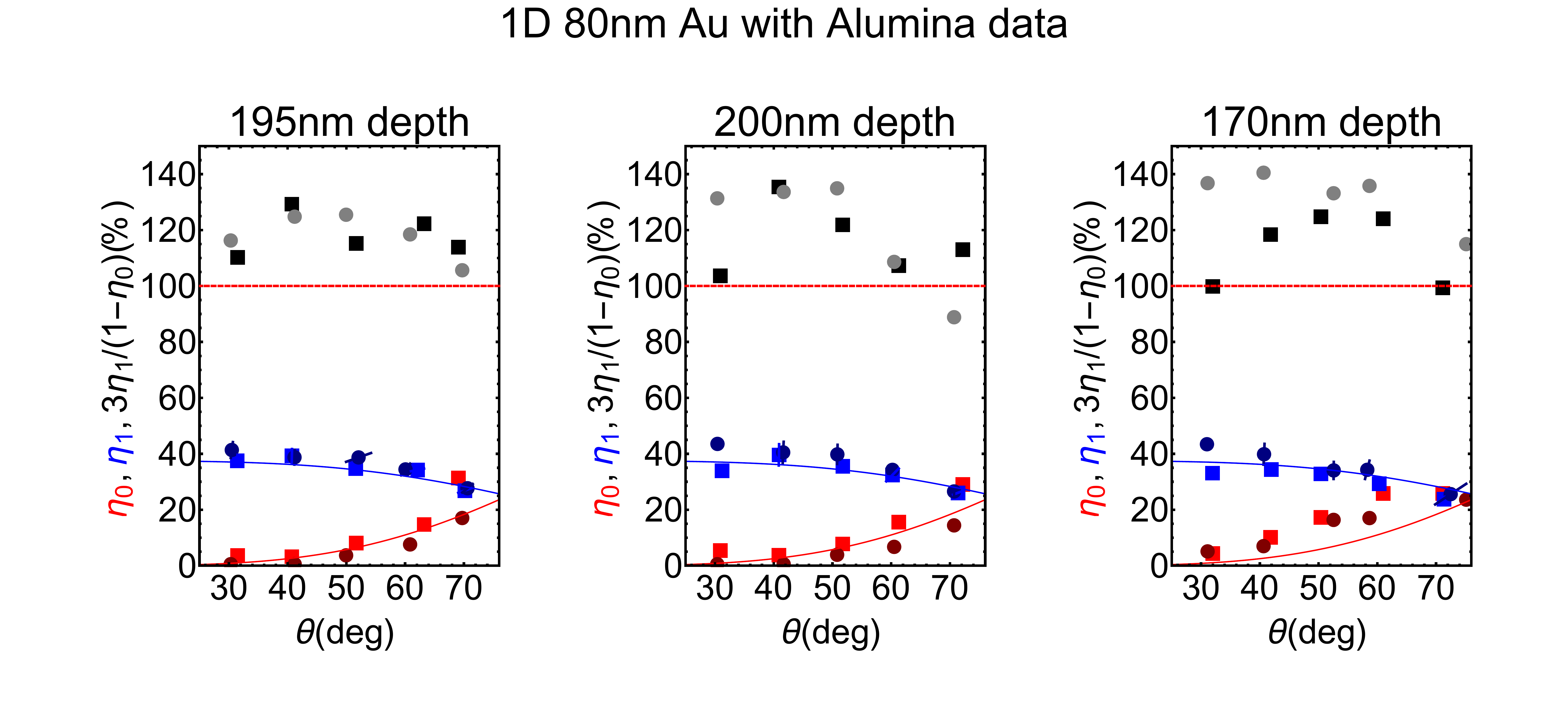}
\begin{minipage}{\columnwidth}
\vspace{2cm}
\end{minipage}
\includegraphics[width=1.8 \columnwidth]{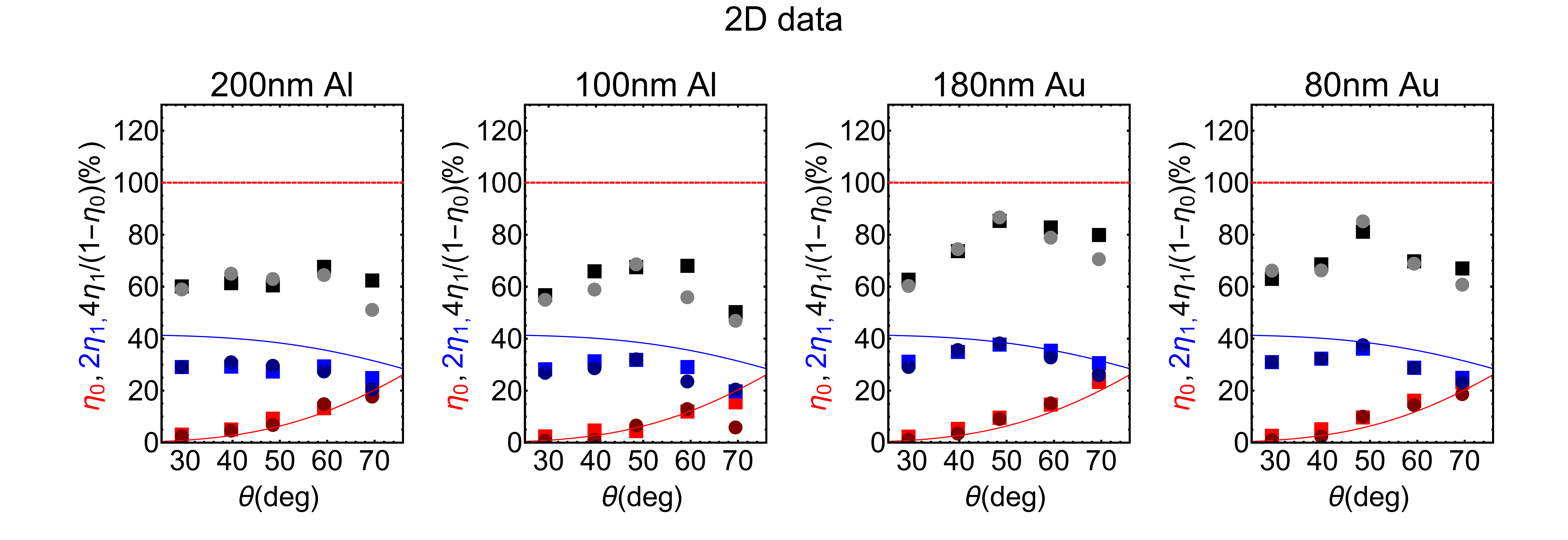}
\caption{Diffraction efficiency for 1D (upper row), 1D alumina coated (middle row) and 2D (lower row) gratings, color scheme as per Fig.~\ref{collective}.}
\label{diff}
\end{figure*}

\begin{figure*}[!ph]
\includegraphics[width=1.8 \columnwidth]{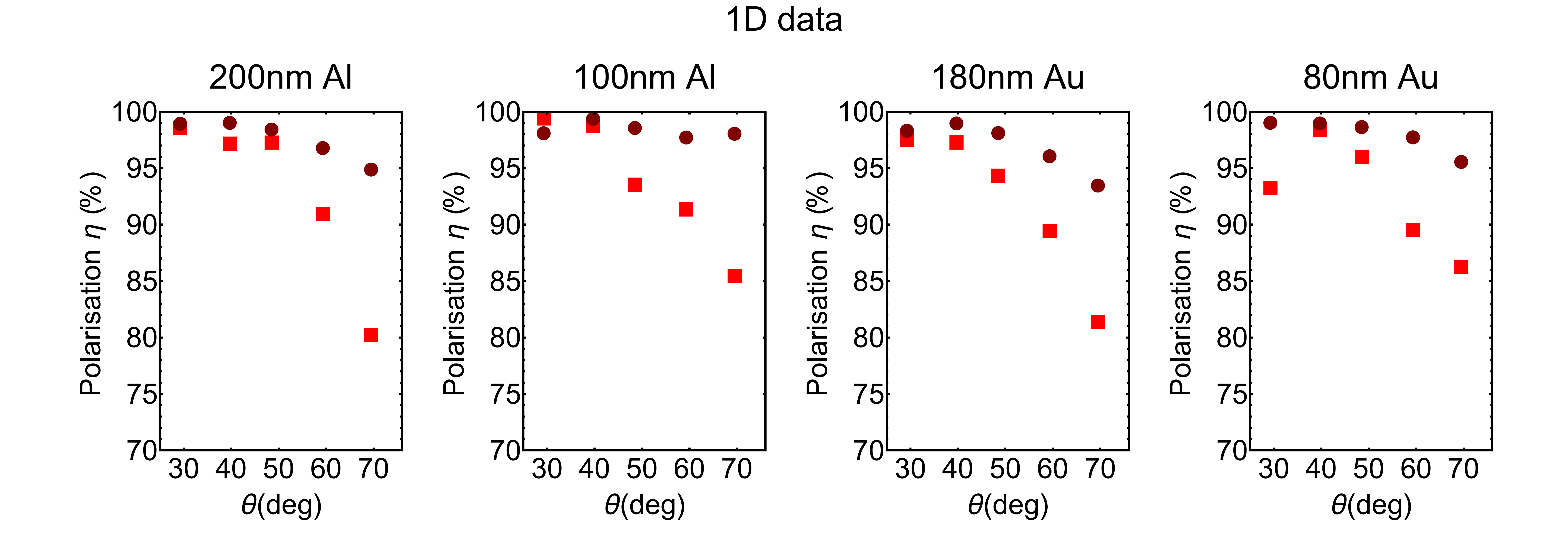}
\begin{minipage}{\columnwidth}
\vspace{2cm}\end{minipage}
\centering
\includegraphics[width=1.45 \columnwidth]{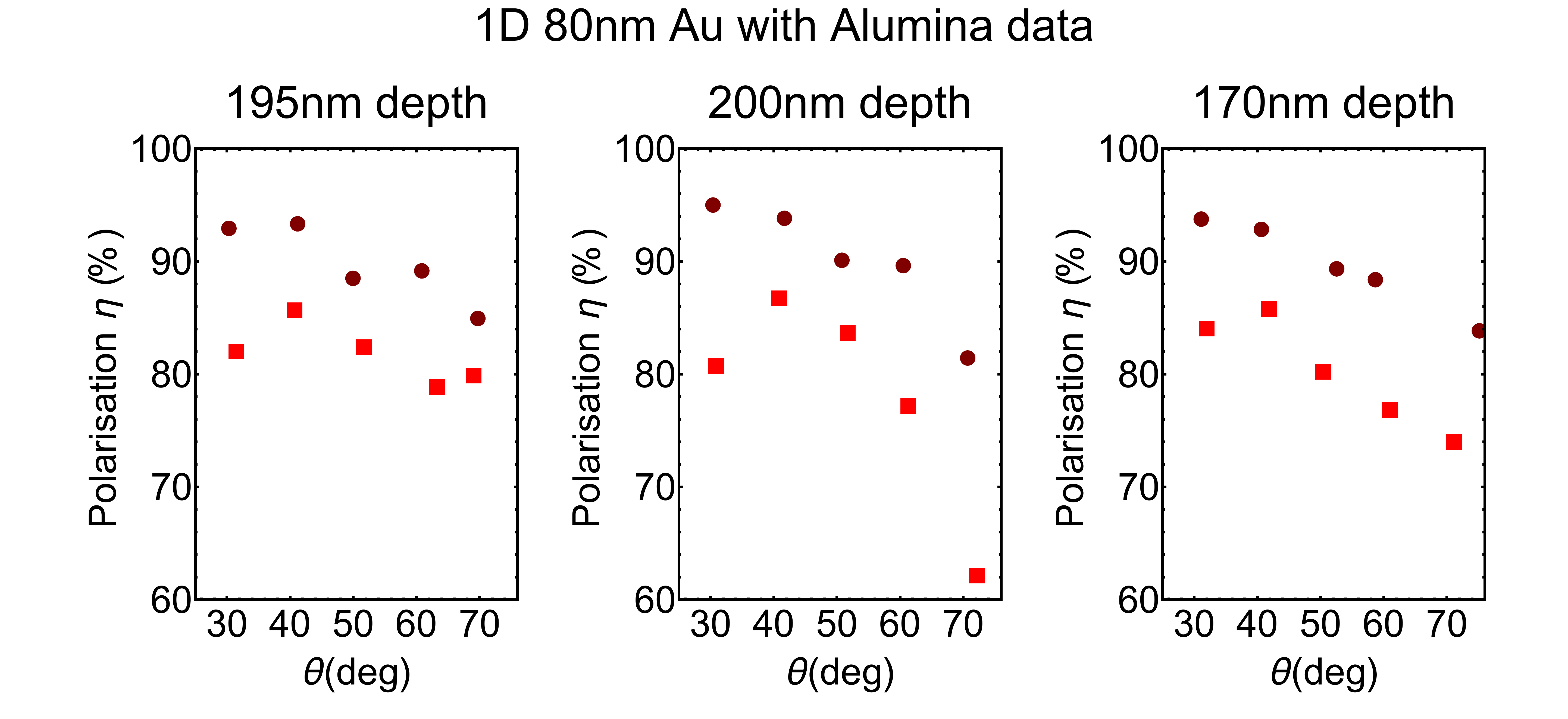}
\begin{minipage}{\columnwidth}
\vspace{2cm}\end{minipage}
\includegraphics[width=1.8 \columnwidth]{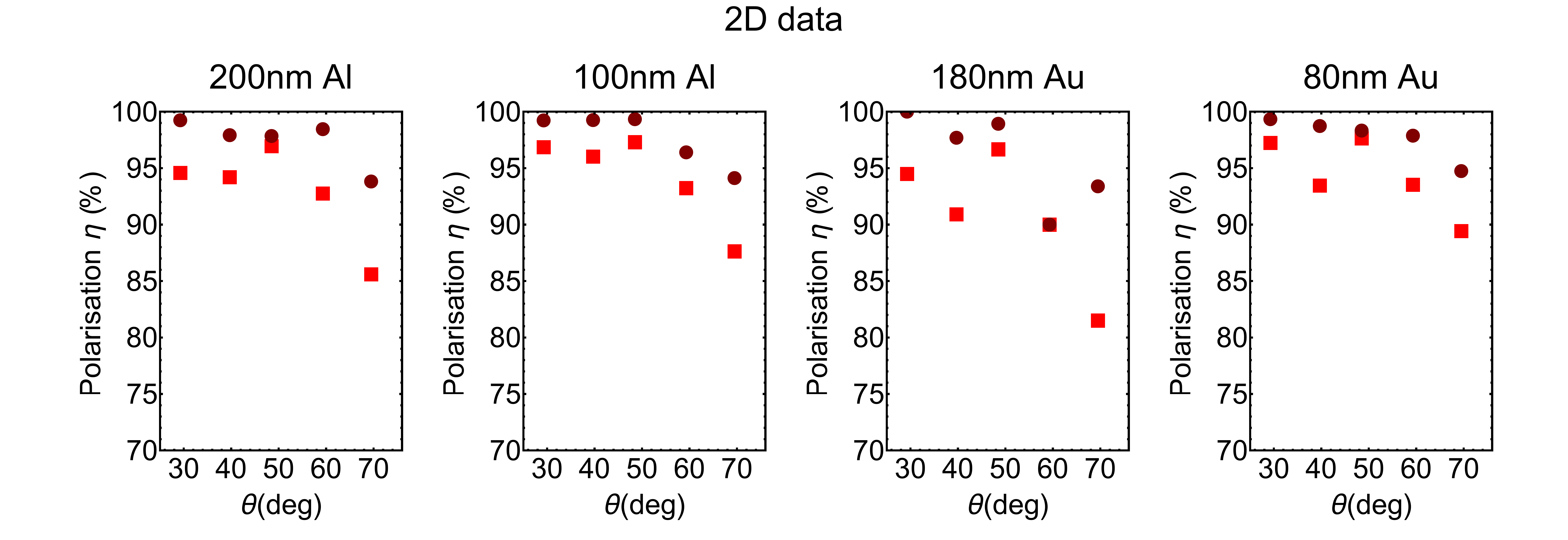}
\caption{Polarisation purity for 1D (upper row), 1D alumina coated (middle row) and 2D (lower row) gratings, color scheme as per Fig.~\ref{polarisation}.}
\label{pol}
\end{figure*}

Please see Figs.~\ref{diff} and \ref{pol} after the references.

\begin{thebibliography}{99}

\bibitem{katori} M.\ Takamoto,  F.~L.\ Hong, R.\ Higashi, and H.\ Katori, ``An optical lattice clock,'' 
\href{http://dx.doi.org/10.1038/nature03541}{Nature {\bf 435}, 321-324 (2005)}.

\bibitem{deutsch}
C. Deutsch, F. Ramirez-Martinez, C. Lacro\^ute, F. Reinhard, T. Schneider, J. N. Fuchs, F. Pi\'echon, F. Lalo\"e, J. Reichel, and P. Rosenbusch, ``Spin self-rephasing and very long coherence times in a trapped atomic ensemble,'' 
\href{http://dx.doi.org/10.1103/PhysRevLett.105.020401}{Phys. Rev. Lett. {\bf 105}, 020401 (2010)}.

\bibitem{poli}
N. Poli, F.~Y.\ Wang, M.~G.\ Tarallo, A.\ Alberti, M.\ Prevedelli and G.~M.\ Tino, ``Precision measurement of gravity with cold atoms in an optical lattice and comparison with a classical gravimeter,'' 
\href{http://dx.doi.org/10.1103/PhysRevLett.106.038501}{Phys.\ Rev.\ Lett.\ {\bf 106}, 038501 (2011)}.

\bibitem{gross}
C. Gross, T. Zibold, E. Nicklas, J. Est\`{e}ve, and M. K. Oberthaler, ``Nonlinear atom interferometer surpasses classical precision limit,'' 
\href{http://dx.doi.org/10.1038/nature08919}{Nature {\bf{464}}, 1165-1169 (2010)}.

\bibitem{golub}
M. A. Golub and A. A. Friesem, ``Effective grating theory for resonance domain surface-relief diffraction gratings,'' 
\href{http://dx.doi.org/10.1364/JOSAA.22.001115}{J. Opt. Soc. Am. A {\bf{22}}, 1115-1126 (2005)}.

\bibitem{loewen}
E. G. Loewen, D. Maystre, R. C. McPhedran and I. Wilson, ``Correlation between Efficiency of Diffraction Gratings and Theoretical Calculations over a Wide Range,'' 
\href{http://dx.doi.org/10.7567/JJAPS.14S1.143}{J. Appl. Phys. {\bf{14}}, 143-152 (1975)}.

\bibitem{bize}
S. Bize, P. Laurent, M. Abgrall, H. Marion, I. Maksimovic, L. Cacciapuoti, J. Grunert, C. Vian, F. Pereira dos Santos, P. Rosenbusch, P. Lemonde, G. Santarelli, P. Wolf, A. Clairon, A. Luiten, M. Tobar, and C. Salomon, ``Cold atom clocks and applications,'' 
\href{http://dx.doi.org/10.1088/0953-4075/38/9/002}{J. Phys. B {\bf 38}, 449-468 (2005)}.

\bibitem{ye}
B. J. Bloom, T. L. Nicholson, J. R. Williams, S. L. Campbell, M. Bishof, X. Zhang, W. Zhang, S. L. Bromley, and J. Ye, ``An optical lattice clock with accuracy and stability at the $10^{-18}$ level,''  
\href{http://dx.doi.org/10.1038/nature12941}{Nature {\bf{506}}, 71-75 (2014)}.

\bibitem{arnold}
M. E. Zawadzki, P. F. Griffin, E. Riis, and A. S. Arnold, ``Spatial interference from well-separated split condensates,'' 
\href{http://dx.doi.org/10.1103/PhysRevA.81.043608}{Phys. Rev. A {\bf 81}, 043608 (2010)}.

\bibitem{kitching}
P. D. D. Schwindt, S. Knappe, V. Shah, L. Hollberg, J. Kitching, L. A. Liew, and J. Moreland, ``Chip-scale atomic magnetometer,''  
\href{http://dx.doi.org/10.1063/1.1839274}{Appl.\ Phys.\ Lett.\ {\bf 85}, 6409-6411 (2004)}.

\bibitem{vangeleyn2}
M. Vangeleyn, P. F. Griffin, E. Riis, and A. S. Arnold, ``Laser cooling with a single laser beam and a planar diffractor,'' 
\href{http://dx.doi.org/10.1364/OL.35.003453}{Opt. Lett. {\bf 35}, 3453-3455 (2010)}.

\bibitem{nshii}
C. C. Nshii, M. Vangeleyn, J. P. Cotter, P. F. Griffin, E. A. Hinds, C. N. Ironside, P. See, A. G. Sinclair, E. Riis, and A. S. Arnold, ``A surface-patterned chip as a strong source of ultracold atoms for quantum technologies,'' 
\href{http://dx.doi.org/10.1038/NNANO.2013.47}{Nature Nanotech.\ {\bf 8}, 321-324 (2013)}.

\bibitem{vangeleyn1}
M.\ Vangeleyn, P.\ F.\ Griffin, E.\ Riis and A.\ S.\ Arnold, 
``Single-laser, one beam, tetrahedral magneto-optical trap,'' 
\href{http://dx.doi.org/10.1364/OE.17.013601}{Opt. Express \textbf{17}, 13601-13608 (2009)}.

\bibitem{chu}
E. L. Raab, M. Prentiss, A. Cable, S. Chu, and D. E. Pritchard, ``Trapping of neutral Sodium atoms with radiation pressure,'' 
\href{http://dx.doi.org/10.1103/PhysRevLett.59.2631}{Phys. Rev. Lett. {\bf 59}, 2631 (1987)}.

\bibitem{monroe}
C. Monroe, W. Swann, H. Robinson, and C. Wieman, ``Very cold trapped atoms in a vapor cell,'' 
\href{http://dx.doi.org/10.1103/PhysRevLett.65.1571}{Phys. Rev. Lett {\bf 65}, 1571 (1990)}.

\bibitem{mcgilligan}
J. P. McGilligan, P. F. Griffin, E. Riis, and A. S. Arnold, ``Phase-space properties of magneto-optical traps utilising micro-fabricated gratings," 
\href{http://dx.doi.org/10.1364/OE.23.008948}{Opt.\ Express 23, 8948-8959 (2015)}.

\bibitem{himsworth}
J.\ A.\ Rushton, M.\ Aldous, and M.\ D.\ Himsworth, 
``The feasibility of a fully miniaturized magneto-optical trap for  ultracold quantum technology,'' 
\href{http://dx.doi.org/10.1063/1.4904066}{Rev.\ Sci.\ Instrum.\ \textbf{85}, 121501 (2014)}.

\bibitem{rolston}
J.\ Lee, J.~A.\ Grover, L.~A.\ Orozco, and S.~L.\ Rolston, 
``Sub-Doppler cooling of neutral atoms in a grating magneto-optical trap,'' 
\href{http://dx.doi.org/10.1364/JOSAB.30.002869}{J.\ Opt.\ Soc.\ Am.\ B \textbf{30}, 2869-2874 (2013)}.

\bibitem{rosenbusch}
R. Szmuk, V. Dugrain, W. Maineult, J. Reichel, and P. Rosenbusch, ``Stability of a trapped atom clock on a chip,''
\href{http://dx.doi.org/10.1103/PhysRevA.92.012106}{Phys. Rev. A {\bf 92}, 012106 (1985)}.

\bibitem{newport}
C. A. Palmer and E. G. Loewen, ``Diffraction grating handbook,''  \href{https://www.google.co.uk/url?sa=t&rct=j&q=&esrc=s&source=web&cd=2&cad=rja&uact=8&ved=0CCkQFjABahUKEwiI4I2EzNPHAhWkmtsKHb9WAKs&url=http%3A%2F%2Fwww.gratinglab.com%2FInformation%2FHandbook%2FHandbook.aspx&ei=BW3kVcjJB6S17ga_rYHYCg&usg=AFQjCNFVMMKttSzZpNs56GR7CuC6dU0M3Q&sig2=S2RhE0OQFMt6CdWD39wLCw}{Springfield, OH: Newport Corporation (2005)}.

\bibitem{tseng}
A. A. Tseng, C. Kuan, C. D. Chen, and K. J. Ma, ``Electron beam lithography in nanoscale fabrication: recent development,'' \href{http://dx.doi.org/10.1109/TEPM.2003.817714}{Electronics Packaging Manufacturing, IEEE transactions {\bf 26}, 141-149 (2003)}.

\bibitem{kley}
U. D. Zeitner, M. Oliva, F. Fuchs, D. Michaelis, T. Benkenstein T. Harzendorf, and E-B Kley, ``High performance diffraction gratings made by e-beam lithography,'' 
\href{http://dx.doi.org/10.1007/s00339-012-7346-z}{Appl. Phys. A {\bf 109}, 789–796 (2012)}.

\bibitem{cotter}
J. P. Cotter \textit{et al.}, submitted (2016).

\bibitem{chu2}
S. Chu, L. Hollberg, J. E. Bjorkholm, A. Cable, and A. Ashkin, 
``Three-dimensional viscous confinement and cooling of atoms by resonance radiation pressure,'' 
\href{http://dx.doi.org/10.1103/PhysRevLett.55.48}{Phys. Rev. Lett. {\bf 55}, 48 (1985)}.

\bibitem{dalibard}
J.\ Dalibard and C.\ Cohen-Tannoudji, 
``Laser cooling below the Doppler limit by polarization gradients: simple theoretical models,'' 
\href{http://dx.doi.org/10.1364/JOSAB.6.002023}{J.\ Opt.\ Soc.\ Am.\ B {\bf 6}, 2023-2045 (1989)}.

\bibitem{NIST}
V. R. Weidner and J. J. Hsia, ``NBS measurement services: spectral reflectance,''  \href{https://www.google.co.uk/url?sa=t&rct=j&q=&esrc=s&source=web&cd=1&cad=rja&uact=8&ved=0CCQQFjAAahUKEwi81pqxy9PHAhUyWtsKHWbuDVg&url=http%3A%2F%2Fwww.nist.gov%2Fcalibrations%2Fupload%2Fsp250-8-2.pdf&ei=V2zkVfzOELK07Qbm3LfABQ&usg=AFQjCNEyCaLqx-XX1tj8Kfn6oN3WKfUjZw&sig2=3V7dWFcusyQ4LMDIlX2pew}
{Center for radiation research, National Measurement Laboratory, NBS, MD 20899 (1987)}.

\bibitem{haring}
T. N. Woods, R. T. Wrigley, III, G. J. Rottman, and R. E. Haring, ``Scattered-light properties of diffraction gratings,'' 
\href{http://dx.doi.org/10.1364/AO.33.004273}{Appl. Opt. {\bf 33}, 4273-4285 (1994)}.

\end{thebibliography}
\end{document}